# Manipulation over Surface Waves in Bilayer Hyperbolic Metasurfaces: Topological Transition and Multidirectional Canalization

Aleksey Girich[1], Liubov Ivzhenko[1,2], Artem Hrinchenko[3], Sergey Tarapov[1], and Oleh Yermakov[3],
*[1]O. Ya. Usikov Institute for Radiophysics and Electronics NAS of Ukraine, Ac. Proskura str. 12, Kharkov, 61085, Ukraine*
*[2]Institute of Spintronics and Quantum Information, Faculty of PhysicsAdam Mickiewicz University in Poznań, Uniwersytetu Poznańskiego 2, 61-614 Poznań, Poland*
*[3]Department of Computer Physics, V. N. Karazin Kharkiv National University, Kharkiv, Ukraine*

***Abstract*—Spoof surface plasmon-polariton is a type of surface wave propagating at the artificially engineered structures in microwave and terahertz ranges. These surface waves are highly important in planar photonic and on-chip devices, integrated circuits, lenses, sensors, and antennas applications. However, it is still a challenge to control the propagation regime of such surface waves including the wavefront shapes and propagation directions. In this letter, we study the surface waves in bilayer hyperbolic metasurfaces and show that interplay between two layers allows to manage their regime of propagation. We demonstrate the switching between angle and number of propagation directions of surface waves at the same frequency. Finally, we demonstrate experimentally the tunable multidirectional in-plane canalization of surface waves by adjusting directions of their propagation with angular range from 0 to 12.8 degrees. The discovered rotation-mediated interlayer coupling of hyperbolic metasurfaces paves way towards efficient in-plane transfer of localized electromagnetic signal.**

## I. Introduction

Metasurfaces have recently gained significant attention due to their ability controlling the incident electromagnetic field on demand, flat geometry, light weight and rich functionality [1]–[3]. They represent the periodic planar arrays of subwavelength resonant elements, usually called *meta-atoms*. In addition to the rich far-field functionality, metasurfaces support the propagation of surface waves (SWs) paving way towards near-field applications such as lensing, antennas, sensing, quantum communications, opto-mechanics, etc [4]–[8]. Even more extraordinary properties are discovered with hyperbolic metasurfaces supporting the propagation of hybrid TE-TM surface waves, named recently as *hyperbolic plasmons* [9]–[11]. Hyperbolic plasmons can exist in the optically homogeneous environment, have a higher localization degree and a larger range of propagation angles versus SWs localized at the interface of bulk hyperbolic medium. Furthermore, they have higher directivity, multiplicity of propagation regimes and polarization states in contrast to conventional 2D graphene plasmons [11]–[15]. It leads to a number of applications and phenomena including negative refraction, enhanced spontaneous emission, hyperlensing and routing of SWs in microwave and optical ranges [16]–[21].

Despite of a plethora of interesting phenomena and applications, related to hyperbolic metasurfaces, the flexible manipulation over surface waves at a specified single frequency is still in a state of research. Some possible ways include the phase- change materials [22], [23], complex excitation sources [24],
[25] and twisted 2D layers [26]–[28]. The latter brings the obvious advantages as far it requires mechanically tunable SWs control instead of the specific materials and sources.

This work proposes a simple way to control the surface waves localized at bilayer hyperbolic metasurfaces. Thus, we exhibit a switching from closed ellipse-like to open hyperbola-like equal frequency contour (EFC), called *Lifshitz-like topological transition* [29], accompanied by a fundamental change in the dynamics of propagating waves. We also show the switching between number of SWs propagation directions from 2 to 4 associated with the evolution of EFC. Finally, we demonstrate the control over the propagation direction of diffractionless SW, so-called *canalization* regime, in a mechanical way, namely by the rotation of the meta-atom. The results obtained form a promising platform for twisted photonics applications, on-demand manipulation over surface waves and spatial dispersion engineering.

## II. Design and Methods

### A. Bilayer Hyperbolic Metasurface Designs

Metasurface is composed of two subwavelength layers separated by the dielectric slab packed in a square array with period $a$ = 6 mm (Figs. 1a-1b). A unit cell of each layer represents a copper H-shaped resonator rotated by the angle $\alpha$ with respect to $y$-axis counterclockwisely (Fig. 1c). The bottom layer resonators are oriented along $y$-axis with $\alpha = 0^\circ$ while the top layers are characterized by $\alpha = 0, 30, 60, 90^\circ$ in four different samples. The metallic resonator is described by the geometrical parameters $l_x = 0.6a$ mm, $l_y = 0.8a$ mm and $t$ = 0.5 mm as it is shown in Fig. 1a. Dielectric slab is a Neltec N9245 laminate with relative permittivity of 2.45 (tangential loss equals 0.0016) and the thickness $H$ = 1.143 mm. Importantly, to satisfy the requirement of rotation the following condition must be fulfilled: $l^2 + l^2 \leq a$.

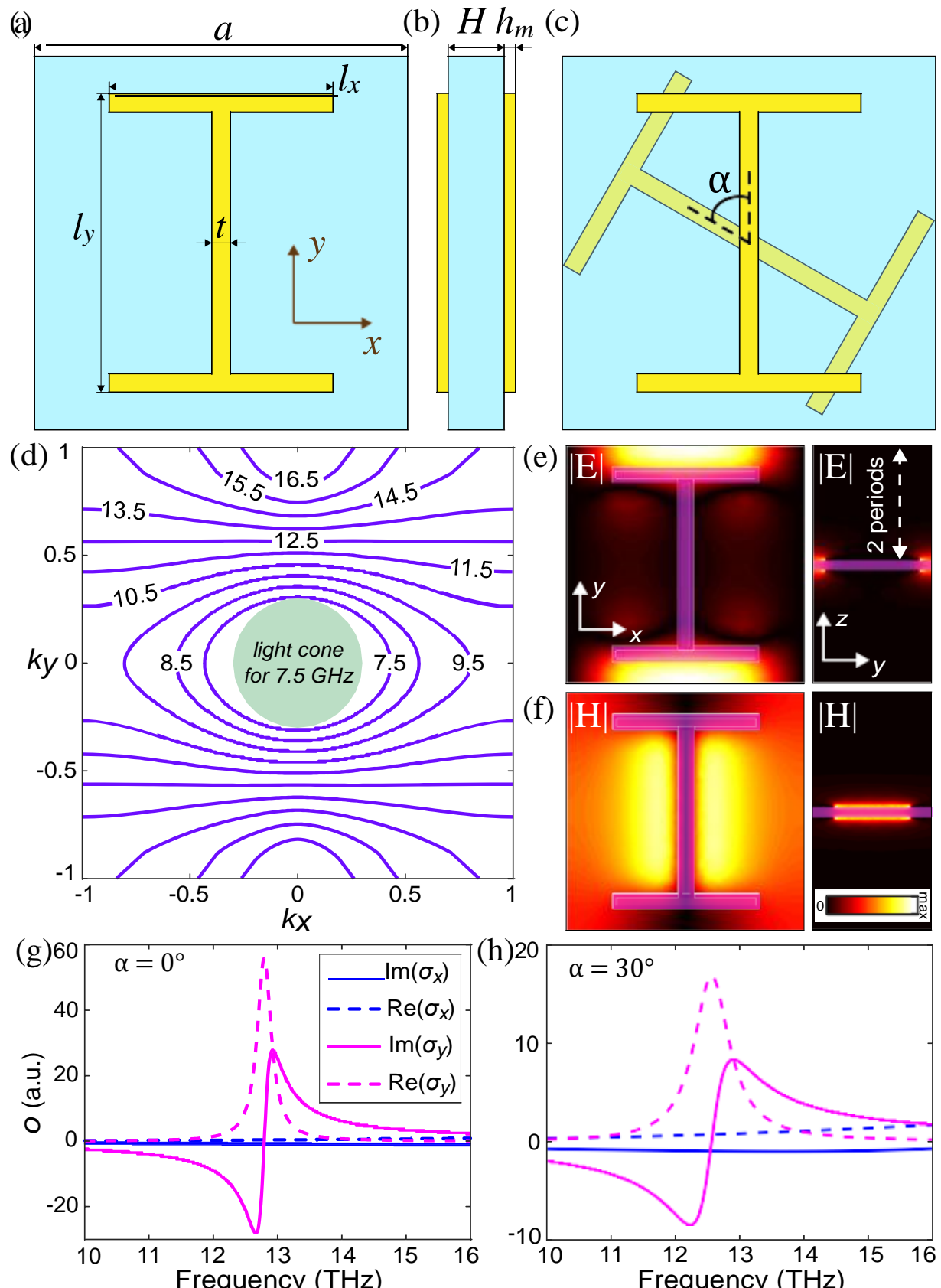

Fig. 1. The geometry of the bilayer metasurface unit cell of period $a = 6$ mm in top (a,c) and side (b) views. Here $\alpha$ is rotation angle between bottom and top copper $H$-elements such as $\alpha = 0^{\circ}$ in (a,b) and $30^{\circ}$ in (c). (d) EFCs of surface waves propagating along the metasurface calculated using CST Eigenmode Solver with periodic boundary conditions for frequencies from 7.5 to 16.5 GHz. Green circle corresponds to the light cone at 7.5 GHz. (e,f) Spatial distributions of electric (e) and magnetic (f) fields of the surface eigenmode at $k_x = \pi/a$ at frequency 9.81 GHz in $xy$ (left column) and $yz$ (right column) planes. The pink regions mark the H-resonator in left column and whole metasurface unit cell in right column. (g,h) The dependence of the effective surface conductivity tensor $x$- (blue lines) and $y$- (magenta lines) components on frequency for (g) $\alpha = 0^{\circ}$ and $\alpha = 30^{\circ}$. The dependecies for $\alpha = 90^{\circ}$ and $\alpha = 60^{\circ}$ are the same as for $\alpha = 0^{\circ}$ and $\alpha = 30^{\circ}$ with mutual $x$-to-$y$ substitution. Here, solid and dashed lines correspond to the imaginary and real parts of the conductivity, respectively. Geometric parameters are the following: $l_x = 3.6$ mm, $l_y = 4.8$ mm, $t = 0.5$ mm, $H = 1.143$ mm, $h_m = 0.035$ mm.

Due to the subwavelength thickness of a metasurface and a negligibly small gap between the copper layers we can effectively consider this bilayer unit- cell structure as an entire meta-atom.

This metasurface is hyperbolic due to the ultra-anisotropic shape of H-resonators and supports hyperbolic spoof surface plasmon-polaritons [21], [30]. The EFCs of SWs supported by the single-layer of this metasurface with $\alpha = 0^{\circ}$ are shown in Fig. 1d. One can see the topological transition from the closed elliptical to open hyperbolic EFCs with the increasing of frequency. The flat EFC at 12.5 GHz corresponds to the difractionless canalization of surface waves characterized by the unidirectional propagation along a line [21], [31]–[33]. Figures 1e and 1d demonstrate the electric and magnetic eigenmode field distributions, respectively, at the boundary of first Brillouin zone ($k_x = \pi/a$) at frequency 9.81 GHz for $\alpha = 0^{\circ}$. The properties of the bilayer metasurface with mutu- ally rotated layers could be described within the transfer matrix formalism with conducting 2D sheets [34]. In Figs. 1g and 1h we show the effective surface conductivity tensors obeying Drude-Lorentz model for the single-layered metasurfaces with different rotation angles $\alpha$ retrieved using the zero-thickness approximation [35]. For $\alpha = 0$ and $30^{\circ}$ the resonant frequency of the $y$-component of surface conductivity is observed in the vicinity of 12.7 GHz, while its $x$-component is near-zero in the frequency range under study. The situation is vice versa for $\alpha = 60$ and $90^{\circ}$ due to the symmetry of the unit cell.

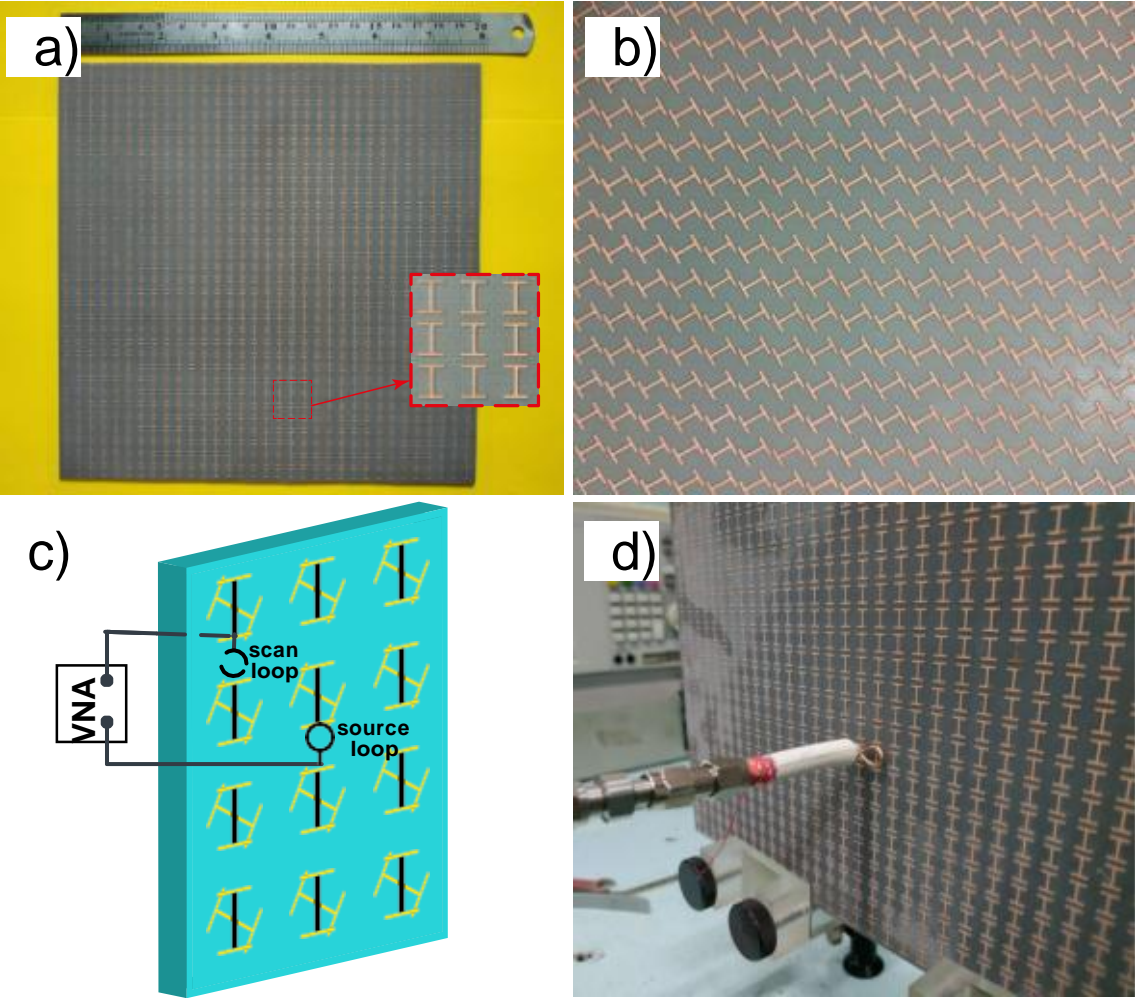

Fig. 2. Photos of the experimental samples with (a) $\alpha = 0^{\circ}$ and (b) $\alpha = 60^{\circ}$. (c) Schematical view and (d) photo of the experimental setup. The source and scan loops are connected to the vector network analyzer (VNA).

### B. *Metasurfaces Fabrication and Measurements*

The four metasurface samples of 33×33 unit cells with $\alpha = 0$, 30, 60 and $90^{\circ}$ were made using the photolithography method (Fig. 2). During measurements the sample was fixed in a vertical position by clamps on a measuring stand. Two loops of the same size served as a source and a receiver of the near-field, respectively. Both loops were located at a distance of 5 mm from the sample (Fig. 2) and connected to Agilent N5230A network analyzer via coaxial lines. The receiver-loop moving in $xy$-plane has scanned the normal component of magnetic field. The scanning area is $100 \times 100$ mm², that is smaller than the size of prototype ($200 \times 200$ mm²). Eventually, we restored the EFCs from the measured field distributions by applying the two-dimensional fast Fourier transformation [36], [37]. Importantly, the most possible excited SW propagation directions could be defined with EFCs as the normals to the contours in the closest points to the centre of k-plane, as it is shown by blue arrows in Figs. 3e-3h, 3m, 3p.

## III. Results Discussion

### A. *Single-Frequency Topological Transition of Surface Waves*

Mechanically tunable bilayer metasurface allows to control flexibly the propagation regime of surface waves including di- rectivity pattern, wavefront shaping and group velocity. Here, we show some SW propagation regimes corresponding to four different metasurfaces under consideration. First, we consider a case of closed EFC at $f = 9.6$ GHz and demonstrate how EFC transforms from elliptical to the rhombus-like shape (Figs. 3a-3h). Starting with $\alpha = 0^\circ$, one can notice the SW propagation along $y$-axis with small divergence (Fig. 3a) that corresponds to the ellipse squeezed along $y$-axis (Fig. 3e). Switching to $\alpha = 30^\circ$ case, the ellipse rotates (Fig. 3f) causing new directions of canalized SWs (Fig. 3b). Further increasing of $\alpha$ leads to the substantial transformation of EFC opening additional SW propagation directions. Thus, at $\alpha = 60^\circ$ the EFC changes to the rounded rhombus-like shape (Fig. 3f) resulting in four channels of SW propagation (Fig. 3c). At $\alpha = 90^\circ$ the rhombus shape of EFC becomes sharper (Fig. 3h) resulting in four-directional canalization of SWs under $\pm 45^\circ$ angle with respect to $y$-axis as it is shown in Fig. 3d.

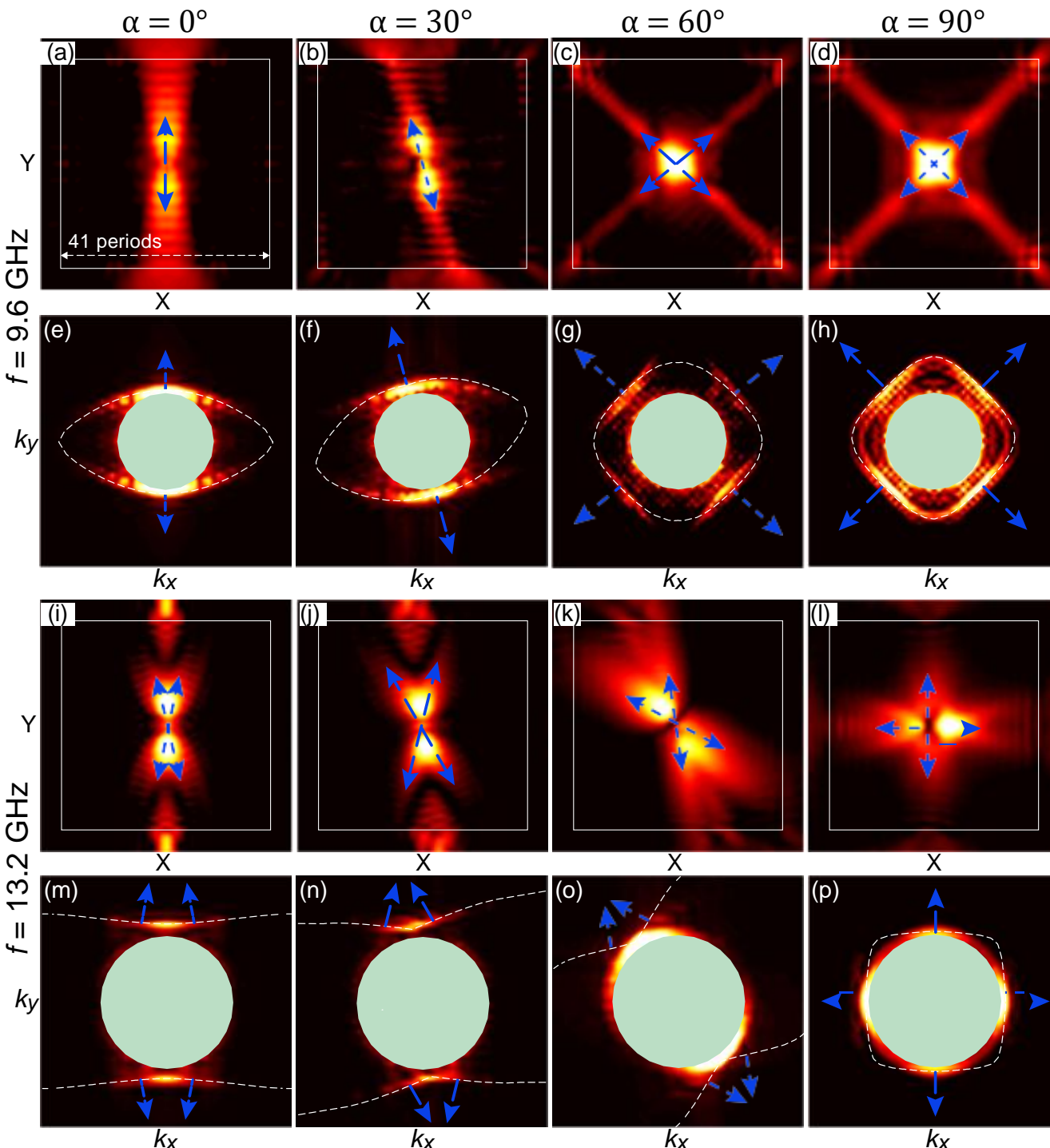

Fig. 3. Regimes of SWs propagation at frequency of 9.6 (a-h) and 13.2 GHz (i-p) with different metasurfaces: (a,e,i,m) $\alpha = 0^\circ$, (b,f,j,n) $\alpha = 30^\circ$, (c,g,k,o) $\alpha = 60^\circ$ and (d,h,l,p) $\alpha = 90^\circ$. (a-d,i-l) The spatial distribution of magnetic field $z$-component excited by the loop located in the center of a metasurface of $41 \times 41$ periods. (e-h,m-p) The EFCs within the first Brillouin zone extracted from the field distributions (a-d,i-l), respectively. White dashed lines show the EFCs calculated theoretically. Green circle corresponds to the light cone, white frame shows the metasurface edges. Blue arrows show the dominant directions of excited SWs propagation according to the simple analysis of theoretically calculated EFCs.

Then, we consider a case of $\alpha$-dependent topological transition of SWs at frequency of 13.2 GHz. One can notice how the EFC changes from the open (Figs. 3m-3o) to closed shape (Fig. 3p). Namely, EFC changes from the almost flat (Fig. 3m) to the hyperbolic ones (Figs. 3n and 3o), and finally to the square-like shape (Fig. 3p). The latter one corresponds to the four quasi-flat wavefronts along $x$ and $y$ axis (Fig. 3l). Another three cases are characterized by the hyperbola-like wavefronts (Figs. 3i-3k) corresponding to the EFCs.

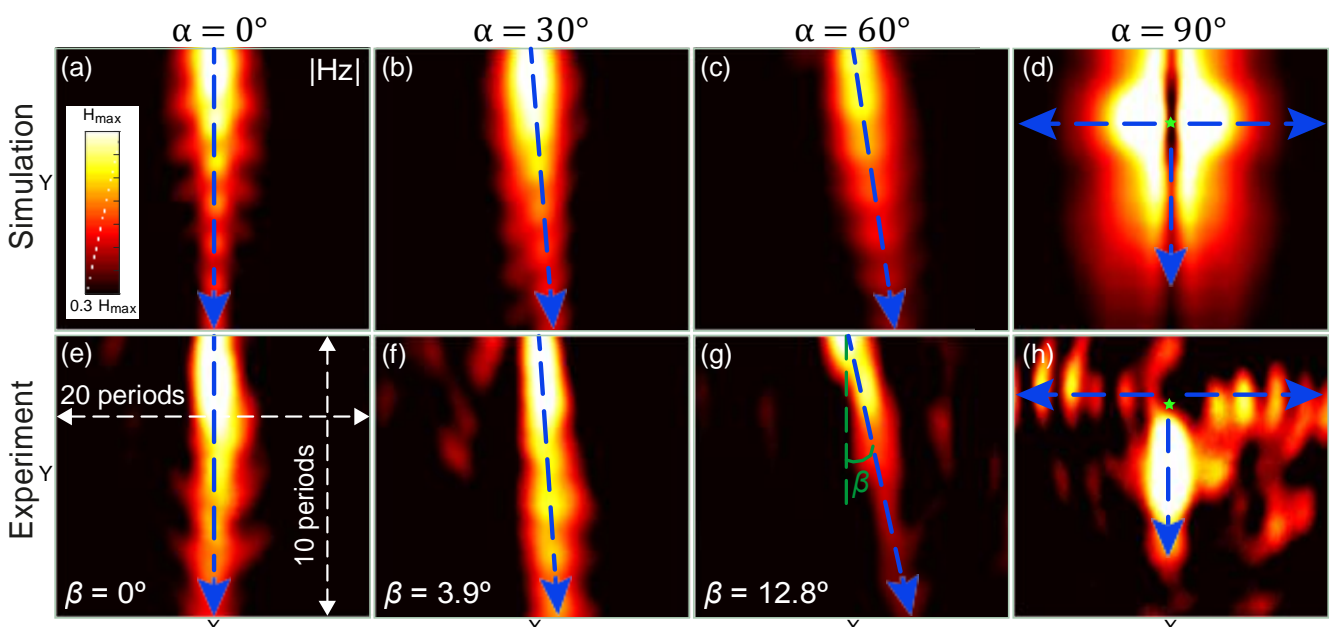

Fig. 4. The spatial distribution of $z$-component of magnetic fields for different metasurfaces (a,e) $\alpha = 0^\circ$, (b,f) $\alpha = 30^\circ$, (c,g) $\alpha = 60^\circ$ and (d,h) $\alpha = 90^\circ$ obtained numerically (a-d) and experimentally (e-h) at frequencies 12.2 and 13 GHz, respectively. Blue arrows show the dominant directions of excited surface waves.

### *B. Multidirectional Canalization of Surface Waves*

Another important issue is the switching of the canalized SWs propagation direction. Canalization being the localized energy transfer along an ultranarrow line is highly important for the planar photonic and on-chip devices, in-plane optical signal control and SW routing from point to point [20], [31]–[33]. One of the challenge is to control the propagation direction of canalized SW. It has been recently achieved for the self-complementary metasurfaces by the direct frequency shift [38], but it is important to implement the switching of SW canalization direction at the single frequency.

Here, we show both numerically and experimentally the canalized SW propagation direction changes between $\beta = 0$, 3.9 and 12.8° for the bilayer hyperbolic metasurfaces under study with $\alpha = 0$, 30 and 60°, respectively, at the operational wavelength $\lambda_\circ = 23.1$ mm (Figs. 4a-4c, 4e-4g). The diffraction angle, propagation length and guided SW wavelength are from 6 to 7° (from 4 to 5.5° in simulations), from 1.6 to 1.8$\lambda_\circ$ (from 3.5 to 3.8$\lambda_\circ$ in simulations) and around $\lambda_\circ/2.5$ ($\lambda_\circ/3$ in simulations), respectively, that is comparable with all known platforms including the interface of wire medium and metal-dielectric layered metamaterial [39], microstructured black phosphorus [31], [33], hyperbolic H-shaped metasurface [21] and twisted van der Waals materials [27]. At $\alpha = 90^\circ$ the canalization takes place in four directions, but less efficiently in comparison with previous cases (Figs. 4d, 4h).

## IV. Conclusion

We have examined the various opportunities of the surface waves control via the mechanical unit-cell rotation of two layers of hyperbolic metasurface in the microwaves. Thus, we demonstrated the control over the wavefront shaping, angle and number of propagation directions at the same frequency. Besides, we have shown the single-frequency topological transition and multidirectional canalization of surface waves. To the best of our knowledge, this work is a first experimental demonstration of the manipulation over the canalized surface wave direction at a single frequency. The concept and methods proposed can be applied for a plethora of far-field and near-field applications within the integrated optical circuits, data processing systems and antenna arrays.

This work was supported in part by NCN OPUS-LAP project no 2020/39/I/ST3/02413 and the Ministry of Education and Science of Ukraine under grant No. 0122U001482. (*Corresponding author: Oleh Yermakov*, oe.yermakov@gmail.com).